# Measuring Global Urban Complexity from the Perspective of Living Structure

Andy Jingqian Xue, Chenyu Huang and Bin Jiang*

LivableCityLAB, Thrust of Urban Governance and Design, Society Hub
The Hong Kong University of Science and Technology (Guangzhou)
Email: andy.j.xue|chuang277@connect.hkust-gz.edu.cn|binjiang@hkust-gz.edu.cn

*(Draft: June 2024, Revise: August, September2024)*

**Abstract**
As urban critic Jane Jacobs conceived, a city is essentially the problem of organized complexity. What underlies the complexity refers to a structural factor, called living structure, which is defined as a mathematical structure composed of hierarchically organized substructures. Through these substructures, the complexity of cities, or equivalent to the livingness of urban space (L), can be measured by the multiplication the number of cities or substructures (S) and their scaling hierarchy (H), indicating that complexity is about both quantity of cities and how well the city is organized hierarchically. In other words, complexity emerges from a hierarchical structure where there are far more small cities or substructures than large ones across all scales, and cities are more or less similar within each individual hierarchy. In this paper, we conduct comprehensive case studies to investigate urban complexity on a global scale using multisource geospatial data. We develop an efficient approach to recursively identifying all natural cities with their inner hotspots worldwide through connected component analysis. To characterize urban complexity, urban space is initially represented as a hierarchy of recursively defined natural cities, and all the cities are then represented as a network for measuring the degree of complexity or livingness of the urban space. The results show the Earth's surface is growing more complex from an economic perspective, and the dynamics of urban complexity are more explicit from nighttime light imagery than from population data. We further discuss the implications in city science, aiming to help create and recreate urban environments that are more resilient and livable by fostering organized complexity from the perspective of living structure.

## 1. Introduction

Sustainable urban planning and design is essentially the problem of organized complexity, as Jacobs (1961) conceived in the classic work *The Death and Life of Great American Cities*. The notion of organized complexity indicates that cities are dealing simultaneously with a sizeable number of factors. But more importantly, all factors are interrelated or organized towards an organic whole, more like a biological entity. The complexity perspective on cities can help better understand the fundamental issues of what cities look like and how cities work and make the urban environments more livable and sustainable (Alexander 1965, Jiang 2019). After three decades of studies, Alexander (2002–2005, 2003) discovered that the living structure is what underlies the notion of organized complexity, and it triggers the sense of wellbeing and beauty from human mind and heart. A living structure is defined as a physical structure consisting of numerous substructures with an inherent scaling hierarchy. Across the hierarchy, there are far more small substructures than large ones; while within each level of the hierarchy, substructures are more or less similar in size. These two properties of hierarchy can be characterized by scaling law (Jiang 2015) and Tobler's law (Tobler 1970), allowing the concept of organized complexity to be quantitatively measured according to the living structure of cities.

Most state-of-the-art complexity theories, originating from traditional sciences such as physics and biology (Shannon 1948, Jacob 1977), are not specifically tailored for urban planning or city science. These theories are limited to understanding the existing complexity in cities, rather than addressing the fundamental issues of creating or designing living structures to make them livable or even more livable.

Moreover, the relationship between urban complexity and human experiences are overlooked, as well the issue of how complexity is influencing wellbeing of urban residents nor the goodness of urban space (Simon 1962, Wolfram 2003). Under organic cosmology, space is neither lifeless nor neutral, but a living structure capable of being more living or less living (Alexander 2002–2005, Whitehead 1929). The notion of livingness is a kind of human experience or a sense of place attachment, synonymous to many well-known concepts such as spatial vitality, order, wholeness and organized complexity. The livingness of space has proved to be a quality that is universally shared among people, regardless of age, gender, or culture. Therefore, the degree of urban complexity, or the livingness of space (L), is quantified by the recursively defined substructures (S) and their scaling hierarchy (H) as $L = S \times H$ (Jiang and de Rijke 2022), according to the underlying living structure. This perspective deviates from conventional theories that describe complexity through the power law (Clauset et al. 2009, Zipf 1949, Mandelbrot 1967) or normal distribution with strict mathematical definitions. Instead, the concept of living structure interprets urban complexity through two fundamental properties or design principles of differentiation and adaptation two meet the two laws above (Alexander 2002–2005, Jiang 2022), which favor statistic over exactitude and recognize the ubiquity of organized complexity on the Earth's surface.

Geographic representations have been dominated by a mechanistic view that treats spaces as collections of discrete places and locations (Descartes 1637, 1954). Conventional representations, such as vector and raster formats, prioritize computational efficiency by representing geographic as simple geometric primitives like points, lines, polygons, and pixels (Goodchild et al. 2007). These elements can precisely reflect geometry, but livingness cannot be fully perceived merely through their geometry (Koffka 1936). For example, the central place theory (CPT) becomes the classical model as it finds out the topology matters in terms of good spatial configuration (Christaller 1933, 1966). Under CPT models, there is an inherent hierarchy of far more small cities than large ones in terms of their connectivity. Thus, a topological representation is proposed (Jiang 2015) to better capture the underlying living structure and its scaling hierarchy of urban environments. In this topology-oriented geographic representation, urban space is depicted as a complex network or semi-lattice (Alexander 1965), where each city is a node, and edges are defined by topological relationships. This perspective views space as a hierarchy—or living structure—of nested subspaces, internally connected to form an organic whole by its inherent scaling of far more small subspaces than large ones. The notion of city here refers to the concept of a natural city, which is defined in a bottom-up manner using geospatial big data such as nighttime light (NTL) imagery and population data. This approach is more effective in revealing the inherent hierarchy of urban space than traditional administrative boundaries (Jiang et al. 2015). However, existing vector-based approaches are computationally inefficient (Jiang et al. 2015, Jiang and de Rijke 2023), making it impractical to conduct spatial livingness mapping and analysis on larger scales, such as at global or continental levels.

The contribution of this paper lies in the organic, dynamic, and global perspective of examining the organized complexity or livingness of cities through quantitative approaches. Specifically, the paper makes the following contributions from three aspects: (1) we develop an efficient approach to recursively delineating all natural cities with their inner hotspots worldwide through connected component analysis, enabling large-scale analysis of urban complexity; (2) we conduct comprehensive case studies to investigate urban complexity on a global scale, using multisource geospatial data spanning over than two decades; and (3) we found that the Earth's surface is growing more complex from an economic perspective, and the dynamics of urban complexity are more explicit from NTL imagery than from population data.

The remainder of this paper is structured as follows. Section 2 introduces the concept of living structure and the two fundamental laws of characterizing living structure, using a typical example from the CPT models. Section 3 presents the efficient approach to delineating natural cities from geospatial big data and calculating the degree of urban complexity from the perspective of living structure under the topological representation. In Section 4, we report on case studies of global urban complexity based on all the natural cities and visualize underlying living structure with its inherent hierarchy in terms of both their geometry and topology. In Section 5, we further discuss the implications of complexity on urban planning and aesthetics. Finally, Section 6 concludes and points to future work.

## 2. The living structure and the two laws for characterizing living structure

Living structure refers to the hierarchical structure consisting of numerous substructures, where both substructures and hierarchy are fundamental to characterizing the complexity. Livingness or organized complexity can be triggered by the underlying living structure of space (Alexander 2002–2005). Thus, the livingness (L) is defined mathematically by the multiplication of the number of substructures (S) and their inherent scaling hierarchy (H). In other words, the more substructures and the higher hierarchical levels a city has, the more complex and living a city is (Jiang and de Rijke 2022). The notion of hierarchy distinguishes complex systems such as physical, biological, societal, and computer systems, from simple ones (Simon 1964). Across the hierarchy, substructures at different scales are differentiated from each other, where there are far more small substructures than large ones. Meanwhile, substructures at the same hierarchical level or same scale are connected to each other, where substructures are more or less similar in size. The two notions about living structure above can be characterized by two fundamental laws: scaling law (Jiang 2015) and Tobler's law (Tobler 1970). To illustrate how complexity emerges from a living structure, we compare three different spatial configurations inspired by CPT following the transporting/traffic principle (K = 4) (Figure 1).

The CPT is the approach to designing spatial configuration of cities based on their quantity, size, and location (Christaller 1933, 1966). The first CPT model (Panel a in Figure 1) has 7 cities (S = 7), each serving an individual unit represented by a hexagon. The second CPT model (Panel b) has 37 cities (S = 37), each also serving an individual unit. The second model is more complex or more living than the first since it has more cities—that is, more substructures in general. However, they both have only one level of hierarchy (H = 1), which means the complexity is 7 × 1 = 7 and 37 × 1 = 37. The third CPT model (Panel c) also has 37 cities, but the 6 cities in green serve all 30 cities in blue, and the largest city in red serves all 6 cities in green (S = 30 + 6 + 1 = 37). Among the 37 cities, there are 3 hierarchical levels (H = 3) distinguished by service areas of units, which means that livingness is 37 × 3 = 111. From the calculation, it is clear that configuration (c) is more organized or more living than both configurations (a) and (b).

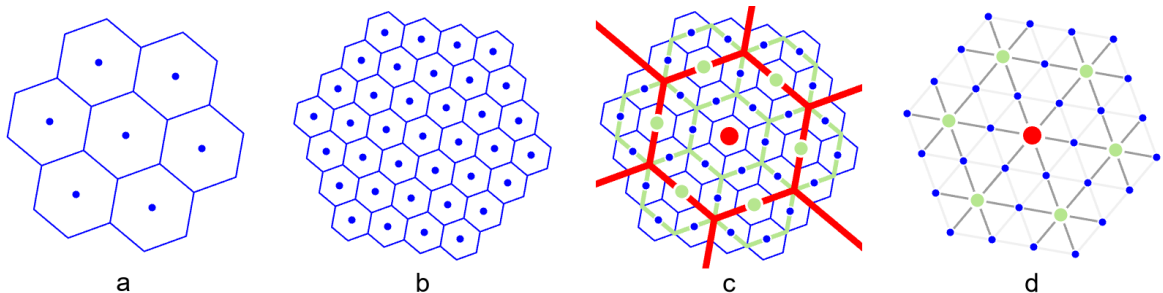

Figure 1: Geometric (a–c) and topological representations (d) of CPT models
(Note: The diagram is the CPT models of K = 4, following the transporting/traffic principle (Christaller 1933, 1963). Model (a) has only 7 cities with only 1 level of hierarchy, model (b) has 37 cities with only 1 level of hierarchy (all in blue), while model (c) has 37 cities with three levels of hierarchy (in red, green, and blue). Thus, model (c) is the most complex, most organized, and most living urban configuration from the geometric perspective. Cities are a coherent whole under the topological representation (d), indicating that cities work and function interconnectedly.)

Taking CPT as an example, let's see how the two fundamental laws can characterize the scaling hierarchy of a living structure. The scaling law indicates that there are far more small cities (30 points in blue) than large ones (7 points in green and red), and the notion recurs twice in the large cities (6 large points in green, and 1 even larger point in red). Yet, cities at each hierarchical level are relatively similar (each serving six lower-level cities) under Tobler's law. These two laws are complementary rather than contradictory to each other in characterizing living structure or organized complexity (Jiang 2022). Statistically, the scaling relationship is reflected as a long tail in a rank-size plot, which can be well described by the power law (Clauset et al. 2009). Conversely, Tobler's law, which is in a bell curve shape according to the Gaussian distribution, states that the closer two substructures are, the more similar they tend to be in size.

It can be clearly seen from the example that cities are living structures from a geometric perspective (Panel a–c in Figure 1). However, we can only understand how these cities look like according to their geometry. Instead, these cities are working or functioning interconnectedly as a coherent whole that each city serves a certain number of lower-level cities. This service relationship forms the basis for the CPT models as a critical reference in urban planning, and livingness comes inherently from their topology. Thus, the topological representation (Panel d) is to build up supporting relationships among individual cities and to discover the underlying scaling hierarchy of urban spaces (Jiang 2019). It is *de facto* a complex network or semilattice defined by directed links (Alexander 1965). Cities at lower hierarchies will point to those at higher levels, indicating that the higher-class cities will serve those lower-class areas. Cities at the same level of hierarchy will point to each other, indicating that they are supporting or serving each other. Cities are a coherent whole under the topological representation and each city has its own degree of wholeness in how it is connected or connected both globally and locally to the surrounding spaces (Alexander 2002–2005, Jiang 2019). The degree of wholeness can be calculated by PageRank score (Page and Brin 1998), which is a variant of eigenvector centrality (Bonacich 1987). The notion of wholeness or PageRank score not only emphasizes not just the number of links a city has, but also the quality of those links. A link from a highly connected city is worth more than a link from a less connected city (see Section 3 for an introduction). Therefore, a structure with a high degree of wholeness is a living structure that reveals organized complexity, while a structure with low wholeness is a non-living or dead structure in contrast.

### 3. The topological representation of cities for computing the degree of complexity
In this section, we introduce the recursive approach to delineating natural cities with their inner hotspots. The topological representation is then introduced for computing the degree of organized complexity or livingness of urban space, taking cities as a coherent whole. Before that, we will illustrate the two geo-referenced datasets we use for identifying natural cities.

### 3.1 Two geospatial datasets
The first dataset is NTL imagery that has been calibrated cross-sensorially using both the Suomi National Polar-orbiting Partnership Visible Infrared Imaging Radiometer Suite (NPP-VIIRS) and the Defense Meteorological Satellite Program Operational Linescan System (DMSP-OLS). The NTL data emulate the NPP-VIIRS standard with a spatial resolution of 15-arc-seconds (approximately 500m at the Equator), covering the period from 2000 to 2022 annually (Chen et al. 2021). Unlike using VIIRS or DMSP data individually, this calibrated NTL combines high spatial resolution with extended temporal coverage starting from 2000 to 2022, enabling a more accurate and comprehensive understanding of human activities on the Earth's surface from a physical and social-economic aspect.

The second dataset is residential population data from the Global Human Settlement Layer (GHSL) project, which depicts the distribution of population in a raster format, with the number of people per cell (Schiavina et al. 2023). This dataset provides residential population estimates between 1975 and 2020 at five-year intervals, along with projections up to 2025 and 2030. The selected population data have a spatial resolution of 1km. As a complement to the NTL data, the population dataset extends the temporal range and offers a human-centric perspective, enhancing our understanding of urban morphology and city structure. For the two datasets, we focus exclusively on continental regions, setting oceanic areas to a *NODATA* value which excludes them from the identification.

### 3.2 The efficient approach to identifying natural cities
The Earth's surface is a living structure composed of natural cities as substructures, and each natural city is a living structure as well comprising inner hotspots recursively. For the sake of convenience, we regard both natural cities and hotspots as natural cities in the following text. Traditionally, a substructure or a natural city of an NTL or population raster is computed to a vector (Jiang et. al 2015, Jiang and de Rijke 2023). However, on the one hand, it is highly time-consuming to conduct recursive approach due to its repetition required in identification of the inner cities. On the other hand, there are large amounts of natural cities at global scale which demands high vacancy of computer memory spaces. Therefore, an efficient approach is proposed that considers a natural city as a raster-based connected component of bright pixels from binary image instead (Samet and Tamminen 1988). As a supplement, Appendix A further details the connected component extraction process and explains why it is more efficient than the conventional approach based on vector.

Pixels lighter than the average value will be considered as natural cities reflecting higher levels of human activity, while darker than the average value will be considered as rural areas. To illustrate, let's take the NTL imagery of New Delhi as an example (Figure 2). The image will be decomposed initially from the whole image to natural cities, and iteratively the natural cities will be decomposed into inner hotspots until there are no more decomposable natural cities nor hotspots, which results in 4 iterations in total (from Panel b1 to b4, Panel d1 to d3). Whether a natural city is decomposable or not depends on the ht-index (Jiang and Yin 2014) calculated by the head/tail breaks (Jiang 2013). The ht-index characterizes the scaling law or the hierarchy. In other words, the larger the ht-index of a city is, the more complex and fractal the city is. One significant finding about scaling law is that the pattern of far more smalls than larges usually recurs more than two times in a living structure. Thus, only if the ht-index of a city is larger than 2, it considered as a living structure that can be decomposed into several substructures (Panel a). At each iteration, only pixels lighter than 0 will be involved in the computation as we only concern the areas that demonstrate human activities. The process is named dichotomization (Panel a1 to a2). After that, the image will be binarized according to the average value (Panel a2 to a3), then each connected component of light pixels will be identified as a natural city or substructure of the entire image (Panel a2 to a3).

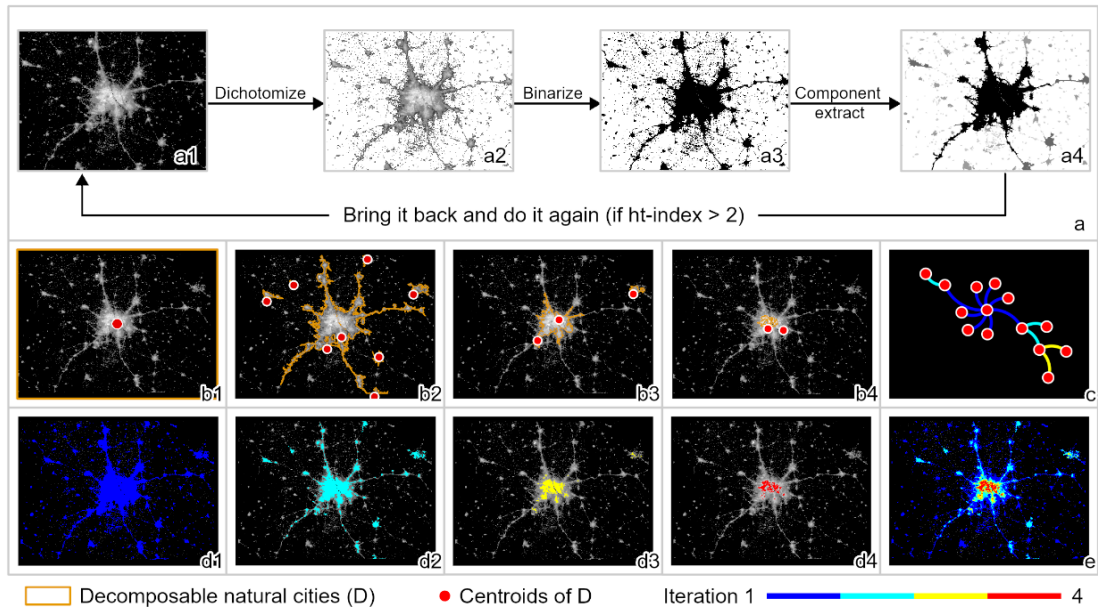

Figure 2: Illustration of the recursive approach to delineating natural cities from NTL imagery (Note: The NTL image of New Delhi is recursively decomposed into numerous natural cities and their inner hotspots. Panel a illustrates how natural cities are derived from each iteration, taking the first iteration as an example. Red points in Panel b1–b4 show the centroids of decomposable substructures or cities at each iteration, with their boundaries in orange. Panel c shows where each decomposable city is originally from. The visualization of natural cities in Panel d1–d4 and Panel e shows cities are a living structure or coherent whole of far more small cities than large ones.)

As shown in Figure 2, starting from the whole image as the largest decomposable substructure (Panel a), it consists of 900x600 (540,000) pixels of which value ranges between 0 and 255. The average value of 190K pixels lighter than 0 is 68. Then they are binarized into 2 groups: 90K urban pixels (47 percent) and 100K non-urban pixels (53 percent). These urban pixels then be identified as connected components of numerous natural cities. The ht-index of these natural cities is 4, indicating the whole image is a decomposable substructure of the living structure then can be derived into further cities. More specifically, the first iteration results in 567 natural cities of which 8 of them are decomposable (Panel b1 to d1). Then the second iteration results in 93 of which 3 of them are decomposable (Panel b2 to d2). The third iteration results in 22 and 2 of them are decomposable (Panel b3 to d3). The fourth and also the last iteration results in 13 natural cities, while none of them are decomposable (Panel b4 to d4). The detailed statistics of Figure 3 are provided in Table 1.

Table 1: Statistics of reclusively derived natural cities and their degrees of livingness
(Note: This table supplements Figure 2 to show the underlying statistics of substructures or natural cities. D = decomposable substructure (natural city), S = substructure (city), U = undecomposable substructure (city), and % = percentage of D to S.)

| Iteration | D | S | % | U |
|---|---|---|---|---|
| 1 | 1 | 567 | 1.4 | 559 |
| 2 | 8 | 93 | 3.2 | 90 |
| 3 | 3 | 22 | 9.0 | 20 |
| 4 | 2 | 13 | 0.0 | 13 |
| ∑ | 14 | 695 | | 3,228 |

Based on the recursively defined natural cities, the hierarchy and differentiation become evident with each iteration. In other words, there are far more low-class cities—which are derived from the early iterations—than high-class ones—which are mainly identified from the late iterations. In the following section, we introduce the topological representations of cities as a coherent whole. These representations unveil differentiation and adaptation within urban systems and provide a definition of the degree of urban complexity from the perspective of living structure and wholeness.

### 3.3 Degree of urban complexity in terms of wholeness

A city is not a tree but a complex network or a semi-lattice where all substructures are interconnected working towards a coherent whole (Alexander 1965). While the recursive approach is able to identify the hierarchy defined by substructures, it overlooks the local adaptation between natural cities. Therefore, topological representation is proposed to construct a directed network of natural cities globally, capturing both differentiation and adaptation among them.

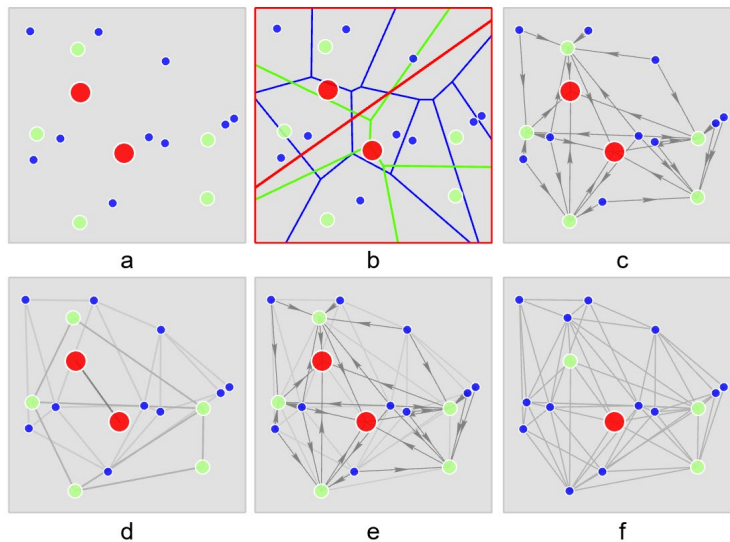

Figure 3: Illustration of constructing topological representation of natural cities
(Note: The 17 fictive cities are derived from 3 iterations (Panel a), indicating 3 hierarchical levels colored by blue, green, and red. The Thiessen polygons are generated based on the centroids (Panel b). Then a complex network is created to capture nested relationships of polygons across the levels (c), and adjacent relationships of the polygons at the same level (c). The scaling hierarchy will be calibrated by their degree of wholeness (from e to f). For example, one of the cities at the highest level belongs to the second level according to its degree of wholeness.)

The topological relationships between cities include containment and adjacency, which both of them can be identified based on the Thiessen polygon (Jiang 2019). Among the hierarchy, the lower-level cities point to the higher-level ones if their Thiessen polygons are intersected. At each level of hierarchy, cities are mutually pointed if their Thiessen polygons are adjacent to each other. As shown in Figure 3, there are 17 cities with a hierarchy of 3 levels or iterations in total (from blue to red in Panel a). By creating Theisen polygon for natural cities at each level (Panel b), there are total 32 edges pointed from lower-level cities to higher-level ones (Panel c). Additionally, there are 35, 14 and 1 pairs of natural cities that are connected locally to among each other from the first iteration to the last one (Panel d).

Compared to metrics like centrality that represents the importance of cities, the PageRank (Page and Brin 1998, Jiang and de Rijke 2022) score better reflects urban space as a coherent whole, capturing local interconnections among cities. PageRank scores are computed iteratively, assessing both the quantity and quality of connections. Starting from an initial score, the iterative computation is based on the principle that links from highly ranked nodes significantly enhance a node's significance more than those from lower-ranked ones. Formally, for each natural city *n*, the PageRank score or the degree of wholeness is defined as follows:

$$PR(n) = \frac{1-d}{N} + d \times \sum_{j \in L(n)} \frac{PR(j)}{C(j)} \quad [1]$$

where *N* is the total amount of natural cities, *C(n)* is the set of all natural cities that link to natural city *n*, *PR(j)* is the current PageRank score of natural city *j*, *C(j)* is the number of outbound links from natural city *j*, and *d* is the damping factor, typically set to 0.85. The influence of a city depends not only on its own connections but also on the importance of other cities to which it is connected. Therefore, the PageRank score can demonstrate the local adaptability between a city and other cities. The interconnections among these cities ultimately build a hierarchy of far more less-connected cities than large ones, thereby facilitating the emergence of complex functions in cities.

From the perspective living structure and topological representations, the degree of urban complexity or livingness of urban spaces (*L*) can be defined by the number of natural cities or substructures (*S*) and their inherent hierarchy (*H*) from wholeness, calculated by head/tail breaks (Jiang 2013):

$$L = S \times H \quad [2]$$

The formula indicates that the more natural cities and the higher levels of hierarchy, the more complex and living the urban spaces are. Based on the topological representation, the degree of urban complexity captures both the differentiation and adaptation among natural cities. For the scenario illustrated in Figure 3, the degree of urban complexity is 17 × 3 = 51. And one red point and one green point are reallocated to its lower hierarchic level due to lack of local adaptation among its surroundings (from Panel e to f of Figure 3).

**4. Case studies from NTL imagery and population data**
This section presents the global results for all recursively defined natural cities by topological representation. The cities are derived from NTL images and population data, and we calculated the degree of urban complexity based on the degree of wholeness or PageRank score for all the supported years by adopted data. The results reveal that (1) the Earth's surface is indeed a living structure or complex system of far more small natural cities than large ones by the topological representation; (2) the spatial configurations of natural cities have grown increasingly complex and living over the past 20 years revealed from NTL imagery (3) the majority of new natural cities have emerged primarily in China, India, and Southeast Asia.

**4.1 Urban morphology revealed from reclusively defined cities**
We applied the recursive approach to identify all natural cities using both NTL and population data. For the sake of better illustration, we selected NTL imagery of 2022 as an example since it covered the latest year and had finest spatial resolution. Figure 4 shows the natural cities and inner hotspots worldwide (Panel 0) with six enlarged regions (Panel 1 to Panel 6). The scaling hierarchy of far more small natural cities than large ones is obvious to be observed visually, thereby the entire Earth's surface is regarded as a living structure. And the geometric boundaries of natural cities and their inner hotspots can be clearly identified based on the NTL imagery of high spatial resolution (approximately 500m). In total, we conducted five iterations to identify all the natural cities.

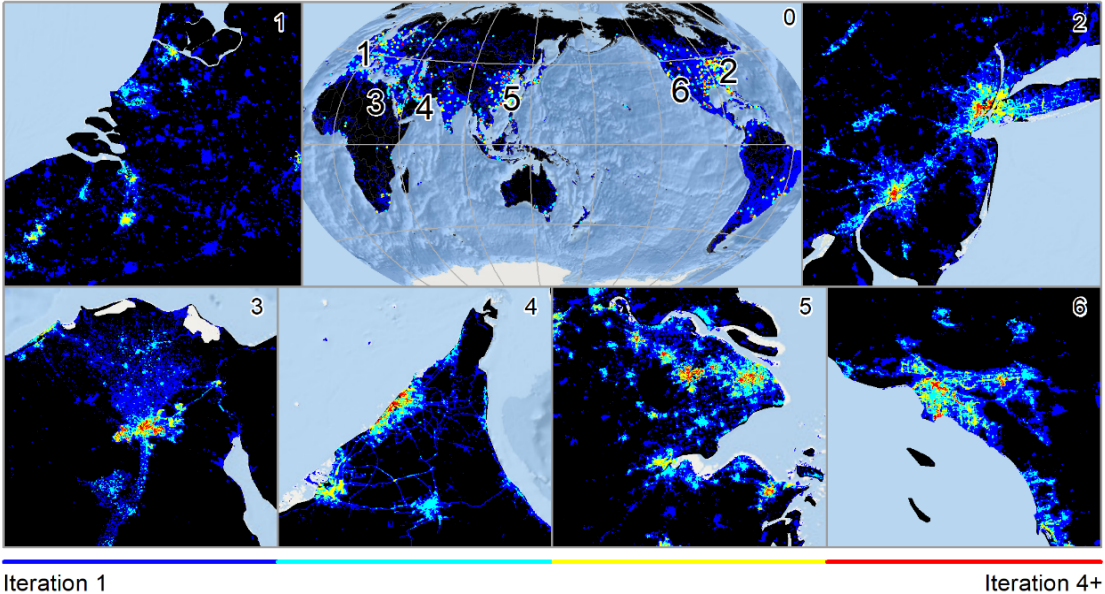

Figure 4: The Earth's surface as a living structure comprising recursively defined natural cities (Note: Natural cities in Figure 4 are delineated from the NTL image of 2022 and cities at each iteration are colored using spectral colormap. Panel 0 presents the specific locations of the six other panels, including Benelux (Panel 1), New York-Philadelphia (Panel 2), Egypt (Panel 3), United Arab Emirates (Panel 4), Yangtze delta (Panel 5), and California (Panel 6). The six regions show the ubiquitous pattern and living structure of far more small cities than large cities on the Earth's surface.)

Table 2: Statistics of natural cities and city hotspots derived from 5 iterations
(Note: This table supplements Figure 4 to show the statistics of recursively derived natural cities, including geometric and social economic aspects. D = Decomposable natural city, S = substructure (natural city derived from D), % = the percentage of S at each iteration relative to the total across all 5 iterations. Average refers to the average area of each S. Average denotes the average area of each natural city. Mean denotes the GDP or population in per unit of pixel.)

| Iteration | Number of cities | | | Area (km$^2$) | | GDP | | Population | |
|---|---|---|---|---|---|---|---|---|---|
| | D | S | % | Sum | Average | Sum | Mean | Sum | Mean |
| 1 | 1 | 173,045 | 85 | 1.23M | 6.5 | 36.4M | 29.7 | 3.33B | 2.72K |
| 2 | 860 | 21,754 | 11 | 203K | 8.0 | 9.19M | 45.2 | 1.21B | 5.94K |
| 3 | 331 | 6,988 | 3 | 41.2K | 4.9 | 2.11M | 51.2 | 391M | 9.49K |
| 4 | 101 | 1,533 | <1 | 5.77K | 3.1 | 303K | 52.7 | 73.4M | 12.7K |
| 5 | 17 | 171 | <0.1 | 430 | 2.1 | 25.1K | 58.5 | 7.02M | 16.3K |
| ∑ | 1,310 | 203,491 | – | 1.48M | – | 48.0M | – | 5.01M | – |

As a supplement to Figure 4, Table 2 presents detailed statistics for natural cities identified at each level of iteration. All natural cities worldwide are derived from the largest decomposable substructure or the whole, which is the NTL imagery of the entire Earth's surface. Approximately 85% of the natural cities are identified at the first iteration, encompassing most of the global area, GDP, and population at the same time. The first iteration results in 173,045 natural cities of which 860 of them are decomposable. Then the second iteration results in 21,754 of which 331 of them are decomposable. The third iteration results in 6,988 and 331 of them are decomposable. The fourth iteration results in 1,533 and 101 of them are decomposable. The fifth iteration, which is also the last iteration results in 17 natural cities, while none of them are decomposable. While the quantity and area of natural cities identified in later iterations are less than those identified earlier, the average GDP and population per unit are higher, indicating that natural cities derived at higher iterations are more socially and economically significant. This suggests that there are far more less-important cities than important ones on the Earth's surface from a socio-economic perspective through the NTL observations.

Compared to the subjectively defined administrative districts set by governments, the concept of natural cities is more advanced in revealing the underlying scaling hierarchy, where there are far more small cities than large ones. The inherent living structure can be clearly seen and making the mapping of global natural cities more living and beautiful, the city centers can be well delineated by the top-level of cities. The topological representation of all the cities can be built from generating the Thiessen polygons (Jiang 2018) and analyzing the containment and adjacent relationships among cities. Thus, the Earth's surface is a coherent whole represented by a complex network. The following section demonstrates the urban complexity calculated from the derived natural cities from NTL images.

**4.2 Urban complexity derived from NTL imagery**
The degree of urban complexity, or the livingness of urban space, can be quantified by examining all natural cities along with their inherent hierarchy. Specifically, the degree of urban complexity (L) can be measured by the product of the number of natural cities (S) and their inherent levels of hierarchy (H). As introduced in Section 3, the notion of hierarchy reflects how cities are differentiated based on their scales, with the topological representation of a complex network and the degree of wholeness reflecting how cities are locally connected to each other.

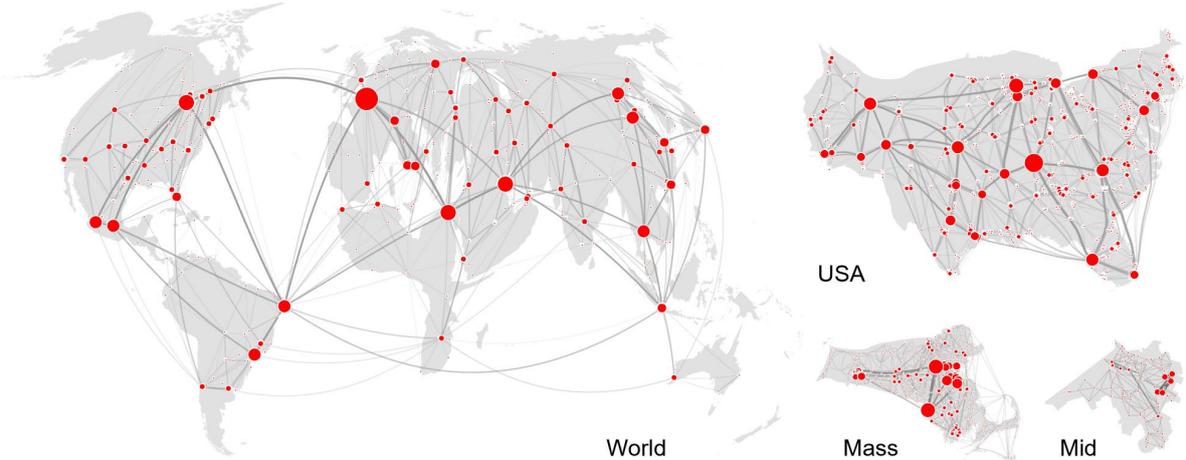

Figure 5: Topological representations of natural cities with their wholeness at global, country, state, and county scale
(Note: The degrees of wholeness for each city are represented by dot sizes. The connections among cities are illustrated by the gray edges. Each panel illustrates the topological presentations of cities in the world, continental USA, Massachusetts (Mass), and Middlesex (Mid), respectively. The background of each representation is the cartogram generated based on its city density.)

In Figure 5, the wholeness of each natural city derived from the NTL imagery of 2022 is visualized by dot sizes. This figure demonstrates that the scaling hierarchy universally exists at the global (Panel a), country (Panel b), state (Panel c), and city scales (Panel d) as well. It shows that there are far more less-connected cities than well-connected ones, indicating that cities indeed form a coherent whole or living structure with an inherent hierarchy. The edges in the topological representation clearly illustrate the interconnections among cities through topological relationships, with cartograms in the background depicting city density. This figure also highlights that some major metropolises, including Paris, Chicago, and Cairo, can be well identified as top-level cities. These cities not only exhibit the highest levels of human activities but are also well-connected to their surrounding cities, providing fundamental and essential services to them. This interconnectedness and service provision further reinforce the notion of a living structure within the urban network. Thus, the topological representation is more effective in showcasing the intricate livingness of space or organized complexity among all cities worldwide from the organic view of space (Jiang 2022).

Table 3: Statistics of urban complexity calculated based on the topological representation
(Note: α represents the power law exponent, xmin is the minimum value of size where the power law behavior is applicable, and p is the goodness-of-fit for the power law model (Clauset et al. 2009). # = number, H = ht-index, and L = degree of urban complexity or livingness of urban space.)

| Regions | # natural cities | # links | | Power law | | | Living structure | |
|---|---|---|---|---|---|---|---|---|
| | | Contained | Adjacent | α | xmin | p | H | L |
| World | 203,320 | 844,076 | 609,705 | 1.9 | 3E-5 | 0.02 | 9 | 2,236,520 |
| USA | 39,701 | 172,958 | 117,951 | 1.9 | 2E-5 | 0.02 | 9 | 357,309 |
| Massachusetts | 522 | 1,673 | 1420 | 2.1 | 7E-4 | 0.00 | 7 | 3,654 |
| Middlesex | 89 | 239 | 196 | 2.3 | 3E-3 | 0.34 | 3 | 267 |

In total, as shown in Table 3, there are over 200K natural cities in the world derived from NTL imagery. The scaling hierarchy can be clearly reflected by the power law exponent of 1.9 with the p value around 0.02 (Clauset et al. 2009, Jiang 2015) and the ht-index of 9. The differentiation from the scaling hierarchy can be observed at the country, state, and city scale as well, with power law exponent around 2 and high ht-index larger than 2. The world has the highest degree of livingness than a country, a state, or a city since it covers all natural cities on the Earth's surface and has the highest levels of hierarchy. This observation is analogous to how the human body, as a complex system, is more complex than any of its individual parts, such as the brain or other organs.

**4.3 Urban complexity derived from population data**
Population data offers a different perspective on urban complexity, which emphasizes more on human presence instead of socio-economic activities. This human-centric view highlights how population distribution shapes urban morphology. Figure 6 illustrates the natural cities in East Africa identified using both NTL imagery (Panels a and b) and population data (Panels c and d), along with their power law statistics of wholeness (Panel e) and sizes (Panel f).

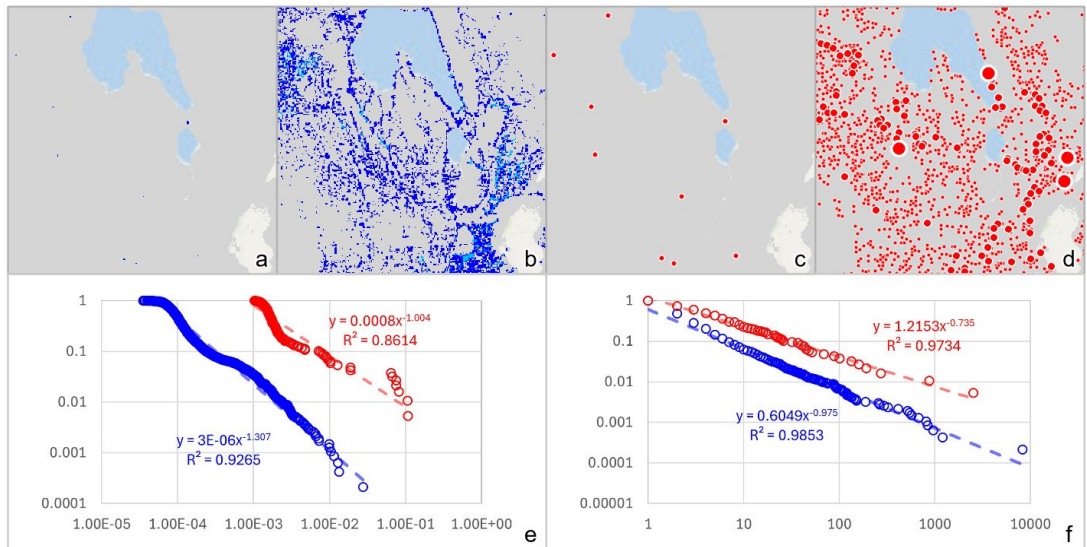

Figure 6: Illustration of natural cities in East Africa derived from NTL imagery and population data, with their power law statistics of wholeness and sizes
(Note: The natural cities derived from NTL and population is shown as Panel a and b, with each city represented by the dots in Panel c and d. Both the distribution of wholeness and city sizes follow power law in Panels e and f.)

The results show there are far more identified cities and hierarchical levels derived from population data, compared to those from NTL imagery. This discrepancy indicates that despite the high population density in the East African region, there are gaps in infrastructure and economic development. The statistics can also reveal the urbanization in East Africa, as the hierarchy of urban development is more clearly seen from the population aspect (Simon 1962). Consequently, relying on a single data source, such as NTL imagery alone, cannot fully capture the intricacies of urban complexity and livingness. This gap underscores the necessity of integrating diverse data sources to achieve a comprehensive understanding of urban dynamics. Our study provides a more complete case analysis by addressing urbanization on a global scale using multi-source data. Table 4 provides more supplemental information for the results in Figure 6.

Table 4: Statistics of urban complexity derived from population and NTL data
(Note: This table supplements Figure 6 to show detailed properties and differences in statistics and urban complexity from the perspective of population data and NTL images.)

| Regions | Data | # natural cities | # links | | Power law | | | Living structure | |
|---|---|---|---|---|---|---|---|---|---|
| | | | Contained | Adjacent | $\alpha$ | xmin | p | H | L |
| Global | Population | 845,979 | 3,839,803 | 2,537,946 | 1.9 | 3E-5 | 0.00 | 7 | 5,921,853 |
| | NTL | 192,279 | 996,504 | 576,561 | 1.9 | 3E-6 | 0.00 | 9 | 1,730,511 |
| USA | Population | 26,283 | 113,264 | 77,596 | 1.9 | 2E-4 | 0.03 | 7 | 183,981 |
| | NTL | 36,914 | 160,198 | 109,609 | 2.0 | 4E-5 | 0.00 | 10 | 369,140 |
| Kenya | Population | 4,704 | 13,732 | 10,006 | 2.5 | 1E-3 | 0.11 | 7 | 32,928 |
| | NTL | 188 | 825 | 504 | 1.8 | 2E-3 | 0.08 | 5 | 940 |

The livingness of urban space reveals large difference between developing (Kenya as an example in Table 4) and developed regions (USA as an example). As shown in Table 4, USA has higher levels of hierarchy derived from NTL imagery than those of Kenya. It indicates that developed countries are more likely to be economically active and have hierarchically organized urban centers. On the other hand, developed countries often display higher degree of livingness in terms of socioeconomics, while developing countries tend to show higher degree of livingness from a population perspective. It is important to note that some p-values are lower than 0.03, suggesting that the distribution of cities' wholeness does not strictly follow a power law. This indicates that the scaling law of a living structure does not fully conform to the precision of mathematical methods like power law, Zipf's law (Clauset, 2009, Zipf, 1949, Mandelbrot, 1967), which cannot fully describe the recursive notion of having far more smalls than larges. Therefore, the ht-index acknowledges the universality of organized complexity and the living structure of cities.

**4.4 City dynamics from the perspective of complexity**
A city as a complex system is changing dynamically with the development of infrastructure and urbanization. Figure 7 illustrates the progressive evolution of natural cities derived from NTL images spanning from 2000 to 2022, presented in the form of kernel density maps which is weighted by the hierarchical levels of each city. The kernel density map portrays the spatial distribution of livingness or organized complexity of cities, with colored areas in red representing densely distributed and high-level cities. Thus, we can not only see the degree of livingness from geospatial data, but can also see where living parts of area are.

The six sub-figures demonstrate the expansion of colored areas over time, culminating in the peak representation in 2022. The comparative development with 2022 is evident, as the sub-figures illustrate the changes in the hierarchical structure represented by different colors. The wider area and more diverse hierarchy of natural cities in 2022, compared to 2000, suggests a substantial transformation, with the establishment of the global pattern of natural cities occurring around the pivotal years of 2010 and 2015. This highlights the significant impact of the 2022 data, emphasizing the substantial changes that occurred from 2000 to 2015. Overall, Figure 7 provides a comprehensive visualization of the transformation and dynamic expansion of natural cities over the years, demonstrating the profound impact of urbanization and human activity on the spatial distribution of NTL. The majority of new natural cities have emerged primarily in China, India, and Southeast Asia. The center of livingness distribution is gradually shifting from developed countries and regions, such as Europe, the USA, and Japan, to developing countries and regions.

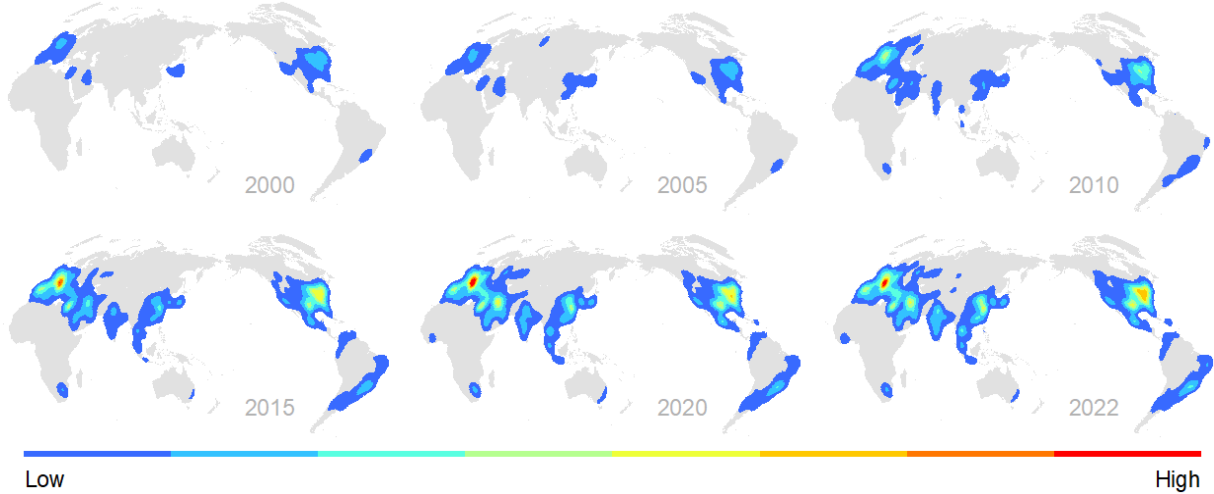

Figure 7: Kernel density maps of natural cities derived globally from NTL images from 2000 to 2022 (Note: The kernel density maps represent the density of natural cities and visualized based on the top 8 classes (9 in total) through head/tail breaks (Jiang 2013) results from the map of 2022.)

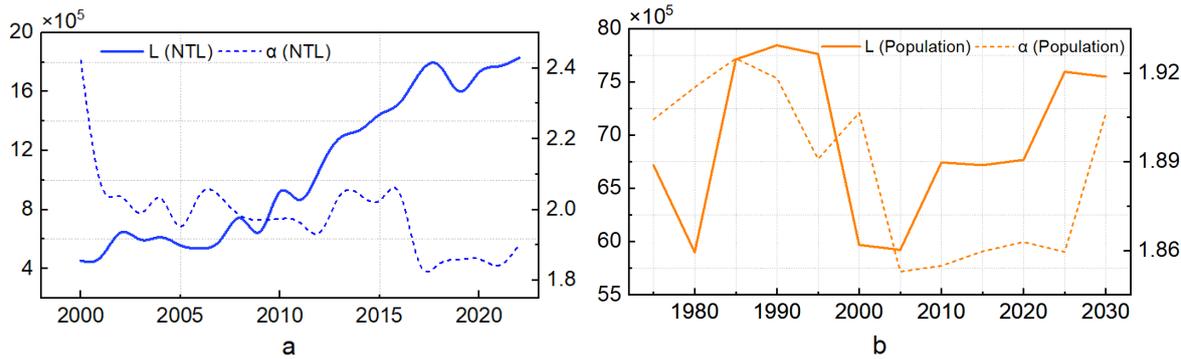

Figure 8: Global urban complexity of each year derived from NTL imagery and population data (Note: Urban complexity is derived from two datasets that present different trends. While the degree of urban complexity shows a significant increase based on NTL imagery, it fluctuates according to population data.)

The illustration in Figure 7 is further supported by Figure 8, which presents the growing trends of global urban complexity validated by the urban complexity or livingness (L) and power-law exponent (α) derived from both NTL imagery and population data. The α value remains stable between 1.8 and 2.1, while the trend in L is more pronounced. Panel a of Figure 8 shows that urban complexity, as calculated from NTL imagery, exhibits a steep upward trend from 2000 to 2022. In contrast, Panel b of Figure 8, which depicts urban complexity calculated from population data, shows more fluctuations, particularly around the year 2000. When comparing Panel b to the reference provided by Panel a, the period following 2000 aligns with the trend revealed by NTL data, thus supporting the reasonable inferences drawn from studies on urban complexity.

## 5. Implications of urban complexity from the new perspective of living structure

The notion of living structure offers a new perspective for measuring organized complexity under the organic worldview (Whitehead 1929). Unlike traditional complexity science represented by sciences (e.g., physics and mathematics), urban planning and design—collectively known as the new city science (Batty 2013) and the new kind of city science (Jiang 2022)—is not only facing the challenge of understanding complexity but also of creating organized complexity in urban spaces, in order to make or remake environments more livable and sustainable. A living structure or design is governed by two fundamental laws: the scaling law and Tobler's law. To achieve this and meet the two laws, urban planners can construct spaces to be more livable or more living by following two key design principles: differentiation and adaptation (Alexander 2002–2005). On one hand, urban space is differentiated into a hierarchy of recursively defined subspaces, where there are far more small cities than large ones. On the other hand, cities are adapted to each other, not only locally but also globally to foster human interaction, community cohesion, and a sense of place (Mahaffy 2017, Lewicka 2011). Therefore, a city as a complex system emerges and evolves under the persistence of differentiation and adaptation towards a coherent whole or living structure. Most of the large urban agglomerations derived as demonstrated in Figure 4 and Figure 5 are effectively organized with their inherent hierarchy by the two laws.

The new perspective from living structure bridges the organized complexity with human perception. Living structure represents the underlying objective factors that evoke a sense of beauty and well-being in human beings. For example, the more living a space is—such as natural scenes—the more healing and nurturing effects it tends to have on people (Ulrich 1984, Wilson 1984). Thus, the notion of living structure has significant applications in computing human perceptions of their surroundings, urban environments, and even artworks or maps (Jiang and de Rijke 2023, Young and Kelly 2017). For example, we can feel the sense of beauty from global natural cites (Figure 4) triggered by its living structure or their inherent scaling hierarchy. The sense of beauty is often said to be in the eye of the beholder, suggesting it varies from one individual to another. However, beauty is actually a matter of fact rather than a subjective opinion or personal preferences (Alexander 2002–2005, Birkhoff 1933, Scruton 2009). Living structure provides a framework to understand the objective and structural factors, expressed through the outer environment of human beings, allowing urban planners and designers to create spaces that are universally perceived as beautiful and harmonious all people, regardless of their age, gender, or culture (Alexander 2002–2005). By following the two fundamental design principles of differentiation and adaptation, planners can design urban environments that are not only functional and efficient but also beautiful or more beautiful, thereby fostering place attachment and healing effects in the hearts and minds of urban dwellers.

The notion of natural cities can uncover the inherent organized complexity of urban systems from a scientific perspective, as demonstrated by the findings of the paper. Natural cities refer to human settlements that emerge organically and objectively from head/tail breaks (Jiang 2013, Jiang et al. 2015). Contrary to conventional administrative boundaries divided by authorities, natural cities are delineated from street nodes by individuals, rather than aggregated population from census data at a collective level. Firstly, the underlying scaling pattern of far more small cities than large ones cannot be well defined within the single and limited range of a country, which determines the vitality of global view of cities that goes across conventional administrative boundaries. Secondly, administrative boundaries strictly follow the top-down principle and are intensively affected by authorities, overemphasizing the impacts of external forces while ignoring the authentic and spontaneous bottom-up thinking implied by natural cities. Thirdly, urban morphology may change over time due to certain spatial facts, e.g., urban sprawl and geographic centroid shift. The reasons above imply that the administrative boundaries tend to be more ambiguous, subjective, and inflexible than natural cities, while natural cities acknowledge the universality of urban studies across the presupposed boundaries, providing the standardized uniformity and possibility of discussing on worldwide urban issues.

The notion of living structure is not intended to demolish the current complex theories, but to offer a supplementary perspective with more consideration of human experiences in a continuous line with the city science (Batty 2013) and the new kind of city science (Jiang 2020). Living structure reflects many key characteristics of complexity, such as holism (e.g. Bohm 1980, Koffka 1936) in contrast to reductionism, hierarchy and the notion of nearly decomposability (e.g. Simon 1962), and complex network or the so-called small world network (e.g. Watts and Strogatz 1998, Barabási and Albert 1999). From the perspective of living structure, a complex system is a coherent whole represented by the topological representation, whose subwholes or substructures are interconnected hierarchically or orderly. Across the hierarchy, substructures are differentiated or heterogeneous, as there are far more small substructures than large ones, while within each level of the hierarchy, they are adapted and homogeneous, being more or less similar in size. Thus, the city as a living structure, is more than the sum of its substructures, with additional functionalities emerging spontaneously through hierarchical self-organization under the holistic or organic worldview (e.g. Whitehead 1929, Ellis 2012, Jiang 2022). More importantly, urban planning can create urban environments where the urban dwellers and the environment form a coherent whole, making the city living more living guided by the two fundamental design principles and mathematical measurement of the livingness of urban space.

## 6. Conclusion

A city is a complex system, with its all the subsystems or subspaces connected internally towards a coherent whole or living structure. The notion of living structure exhibits an inherent hierarchy governed by two fundamental laws: scaling law across hierarchy where there are far more small substructures than large ones, and Tobler's law at each hierarchical level where substructures are more or less similar in size. In this paper, we adopted the notion of living structure as a new perspective and approach for measuring the degree of organized complexity. Through NTL imagery and population data, we found the Earth's surface is a living structure that reveals a high degree of organized complexity or livingneess, which is calculated based on the topological representation taking cities as a coherent whole. To recursively delineate all natural cities on a global scale, we developed an enhanced approach based on connected component extraction, making large-scale analysis on urban complexity feasible and more efficient. Through cases studies on a global scale, we also found NTL imagery and population data unveil different results from each other, while the former one is characterizing more on social-economic effects rather than human presence. The results showed some developing countries might have been urbanized in terms of the population, while it is not urbanized in terms of economics. The spatial heterogeneity is more obvious to be seen from the NTL imagery since it has higher levels of hierarchy or higher ht-index. The spatial configurations of natural cities have grown increasingly complex over the last 20 years revealed from NTL imagery, and the majority of new natural cities have emerged primarily in developing regions including China, India, and Southeast Asia.

The concept of living structure offers a unique perspective, not only for understanding complexity in urban spaces but also for creating organized complexity that makes cities more livable and sustainable. This new kind of urban complexity can be objectively computed to guide the two fundamental design principles of differentiation and adaptation. Urban environments can be built to be more resilient, with cities differentiated to form hierarchies and adapted to each other, both locally and globally. The notion of organized complexity is also seen as a physical phenomenon that triggers human senses of beauty and belonging from the perspective of living structure. With the flourishing development of geographic big data, the livingness of geographic spaces can be better perceived and explored. This means that big data and living structures not only provide new forms of data but also offer a new paradigm and thinking, allowing us to understand space from various scales and dimensions, particularly with a human-centered approach. Beyond the NLT data and population data used in this study, street view images, biometric data, and social media data can also help in quantitatively observing and understanding human perceptions of the environment, ultimately assisting in the urban design and making a living or even more living city.

**Data and code availability statement**
The NTL images (Chen et al. 2021) and population data (Schiavina et al. 2023) used in this study are publicly available at the following links: https://doi.org/10.7910/DVN/YGIVCD and https://human-settlement.emergency.copernicus.eu/ghs_pop2023.php, respectively. The recursive approach to deriving natural cities and calculating the urban complexity is implemented through Python 3.11. The demo example of deriving natural cities from NTL global imagery can be accessed at GitHub through https://github.com/AndyXue957/NaturalCitesfromGlobalNTL. The recursively defined natural cities of each year are in *geotiff* format are available at https://doi.org/10.6084/m9.figshare.25981954.v2. The software tool used in this study includes ArcGIS Pro 3.0.1 for processing and visualizing geospatial data and Microsoft Excel for storing and analyzing numerical data.

**Acknowledgement**
This paper is an extension of a course report for the Urban Complexity Analysis and Modelling course,

taught by Bin Jiang at The Hong Kong University of Science and Technology (Guangzhou). We would like to express our gratitude to the two anonymous referees for their constructive feedback, which has greatly enhanced the quality of the paper.

**Appendix A: A natural city as a connected component of pixels**

A natural city is identified as the brightest hotspot region in NTL imagery and is technically defined as a connected component of a cluster of pixels. Using a recursive approach (as introduced in Section 3) involves repeatedly delineating inner hotspots within these larger natural cities. On the one hand, it is highly time-consuming due to the repeated identification of inner hotspots. On the other hand, the large number of natural cities globally demands significant computer memory space. The conventional approach to identifying natural cities from NTL imagery involves vectorizing binary images, with bright pixels representing urban areas while dark pixels representing rural areas (Jiang et al., 2015). This method converts raster data into vector polygons composed of numerous points and edges (Jiang and de Rijke, 2023). Although vector format provides more details about boundary than raster, the vectorization wastes large amounts of memory spaces and increases the computing time. To enhance the efficiency of identifying natural cities on a global scale and reduce computer memory usage, the new recursive approach treats a natural city as a connected component of pixels instead of a vector polygon (Figure A1).

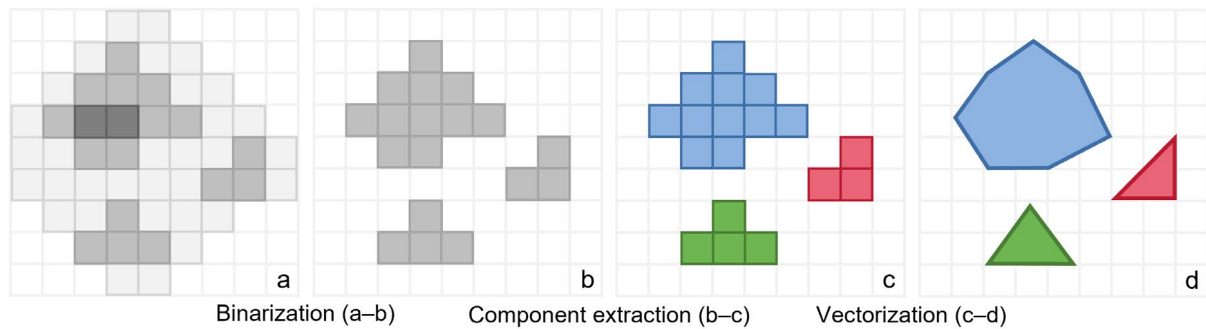

Figure A1: Illustration of raster-based and vector-based approach to delineating natural cities
(Note: In order to identify natural cities, the NTL imagery (Panel a) is binarized by the average value of brightness, where darker pixels represent urban areas, and lighter pixels represent rural areas (Panel b). By connected component extraction, three different regions of dark pixels are identified maintaining the raster format (Panel c). The three colors represent three distinct natural cities. The traditional approach (Jiang and de Rijke 2023) further converts the connected regions into vectors (Panel d), thereby the process is much slower than the new implementation.)

Connected component extraction clusters a set of connected pixels from binary images (Samet and Tamminen, 1988). It labels all the pixels in a detected component with the same value to identify different clusters or components. The process of generating connected components maintains the raster format of natural cities, aligning with the structure of NTL imagery. The new approach is notably more computationally efficient compared to vectorizing natural cities. There is no need to store a separate vector file for each natural city, resulting in significant savings in computer storage space. For example, after connected component extraction, the detection of three natural cities (Panel b) results in their representation as three distinct pixel-based connected regions (Panel c), rather than converting them into continuous areas represented by polygons (Panel d). By assigning three distinct values, the connected component analysis can successfully delineate different natural cities from NTL imagery without converting them into different formats like vector polygons. This ensures significantly higher computational efficiency and reduces space occupation compared to the conventional approach, resulting in at least a tenfold reduction in processing time.